\documentclass{article}

\usepackage{arxiv}

\usepackage[utf8]{inputenc} 
\usepackage[T1]{fontenc}    
\usepackage{hyperref}       
\usepackage{url}            
\usepackage{booktabs}       
\usepackage{amsfonts}       
\usepackage{nicefrac}       
\usepackage{microtype}      
\usepackage{lipsum}

\title{HLA predictions from long sequence read alignments, streamed directly into HLAminer}

\author{
   Ren\'e L.~Warren\\
  Genome Sciences Centre, BC Cancer\\
  Vancouver, BC, V5Z 4S6, Canada \\
  \texttt{rwarren@bcgsc.ca} \\
}

\begin{document}
\maketitle

\begin{abstract}
The rapidly changing landscape of sequencing technologies brings new opportunities to genomics research. Longer sequence reads and higher sequence throughput coupled with ever-improving base accuracy and decreasing per-base cost is now making long reads suitable for analyzing polymorphic regions of the human genome, such as those of the human leucocyte antigen (HLA) gene complex. Here I present a simple protocol for predicting HLA signatures from whole genome shotgun (WGS) long sequencing reads, by directly streaming sequence alignments into HLAminer. The method is as simple as running minimap2, it scales with the number of sequences to align, and can be used with any read aligner capable of sam format output without the need to store bulky alignment files to disk. I show how the predictions are robust even with older and less [base] accurate WGS nanopore datasets and relatively low (10X) sequence coverage and present a step-by-step protocol to predict HLA class I and II genes from the long reads of modern, third-generation sequencing technologies.  

\textbf{Availability:} HLAminer is released under the BC Cancer software license agreement (academic use) and is publicly available from \url{https://github.com/bcgsc/HLAminer}.
\end{abstract}

\keywords{HLA \and HLA typing \and long sequencing reads \and whole genome shotgun sequencing \and HLAminer \and Nanopore \and PacBio}

\section{Introduction}
The landscape of commercially available nucleic acid sequencing technologies is rapidly changing\cite{boti}, and in recent years, long (N50 $\sim$ 10-20 kbp) and ultra-long (<1 Mbp) DNA sequencing reads such as those offered by Pacific Biosciences of California, Inc. (PacBio, Menlo Park, CA) and Oxford Nanopore Technologies PLC (ONT, Oxford, UK) are becoming embedded in more genome sequencing projects\cite{wenger}\cite{cherry}. This is in part because long reads are able to span large genomic repeats, enable allele phasing, and even provide methylation signals without the need for specialized nor additional sample preparation\cite{akbari}. And with the recently released Q20+ chemistry from ONT and CCS (HiFi) technology from PacBio, the base accuracy gap between long and short reads is shrinking, providing unprecedented opportunities for the analysis of genomes, including the characterization of challenging polymorphic gene loci like the Major Histocompatibility Complex (MHC).

The human MHC, also known as the Human Leucocyte Antigen (HLA) complex, is a multi-gene locus with closely-linked variable genes, which play key roles in adaptive immunity. HLA includes two main classes of genes (I and II), which are further divided into several main groups (class I: A, B, C; class II: DQA1, DQB1, DRB1, DPA1, DPB1). Overall, thousands of genes have been catalogued\cite{robinson}, each one encoding a version of the HLA cell surface receptor with a certain specificity to foreign-derived antigen peptide epitopes. HLA-epitope presentation to T cells is at the heart of the immune response in humans, pinning HLA as a disease determinant\cite{dendrou}. HLA can also inform disease susceptibility/risk in population/patient cohorts\cite{warren2} and in the clinic, knowing the HLA signature of patients is vital to organ transplant, to ensure a match between graft and host, for instance.   

A decade ago, I developed HLAminer, the first bioinformatics approach for automated prediction of HLA genes from random whole-genome shotgun (WGS), transcriptome (RNA-seq) or whole-exome sequencing (WES) short read datasets, alleviating the need for additional sample preparation, sequencing kit, costs, specialized and manual labour typically required for analysis of clinical HLA typing data\cite{warren}. The approach originally relied on either targeted \textit{de novo} sequence assembly or direct sequence alignments of short, Illumina-sized reads. Over time, other methods followed suit and new approaches emerged\cite{optitype}\cite{kourami}. Back in 2018, I adapted HLAminer (v1.4) to predict HLA genes from any long sequences, by streaming sequence alignment output directly into HLAminer. Here I present a simple protocol along with benchmarks using four ONT PromethION and PacBio HiFi whole genome shotgun datasets derived from three individuals with known, clinically-derived HLA types.  
\par

\section{Methods}
\subsection*{Data}
Nanopore long read WGS datasets for human individuals NA12878 and NA19240 and PacBio CCS data for individual NA24385 were downloaded from the ENA \url{https://www.ebi.ac.uk/ena} using accessions SRR10965087, ERR2585115 and SRX5327410, respectively. The NA24385 (HG002) ONT dataset was downloaded as per the instructions provided in  \url{https://labs.epi2me.io/gm24385_2021.05/}. HLAminer requires up-to-date HLA genes files (Protocol section step 2, below) and, to be successful, predictions from direct read alignments must include all known HLA genes and unrelated genomic regions in one reference (GCA\_000001405.15\_GRCh38\_genomic.chr-only-noChr6-HLA-I\_II\_GEN.fa.gz), to help prevent spurious off-target long read alignments to HLA genes.
\par

\subsection*{Runs}
HLAminer (v1.4, commit:c00effa) ran on each dataset using a dedicated server-class system with 144 core Intel(R) Xeon(R) Gold 6150 CPU @ 2.70GHz, using a single command combining minimap2\cite{heng} (tested with versions 2.12-r827 and 2.20-r1061) and HLAminer\cite{warren} via a unix pipe (|), and specifying the minimap2 alignment presets (-ax map-ont or map-hifi for ONT or PacBio datasets, respectively) and the MD tag output (e.g. minimap2 -t 48 -ax map-ont --MD GCA\_000001405.15\_GRCh38\_genomic.chr-only-noChr6-HLA-I\_II\_GEN.fa.gz YOUR-LONG-READS.fq.gz | HLAminer.pl -h HLA-I\_II\_GEN.fasta -s 500 -q 1 -i 1 -p hla\_nom\_p.txt -a stream). 
\par

\subsection*{Protocol}
\textbf{1. Clone the HLAminer github repository}\par
\hspace{0.3cm}git clone https://github.com/bcgsc/HLAminer/\par
\textbf{2. Download current HLA genes and update reference files}\par
\hspace{0.3cm}   cd HLAminer/HLAminer-1.4/database\par
\hspace{0.3cm}   ./updateAll.sh\par
\hspace{0.6cm}   -or- \par
\hspace{0.3cm}   wget https://www.bcgsc.ca/downloads/btl/hlaminer/GCA\_000001405.15\_GRCh38\_genomic.chr-only-noChr6-HLA-I\_II\_GEN.fa.gz
\par
\textbf{3. Run HLAminer}\par
\hspace{0.3cm}    cd ../test-demo/\par
\hspace{0.3cm}    ./HPRAwgs\_ONTclassI-IIdemo.sh\par
Note: more general shell scripts are provided with the HLAminer v1.4 distribution (cd bin), for predictions from FASTA/FASTQ (gzip or not) WGS (HPRAwgs\_ONTclassI-II.sh) or RNA-seq (HPRArnaseq\_ONTclassI-II.sh) nanopore long read datasets. For PacBio reads, simply change the minimap2 -ax alignment preset in those files.

\section{Results}
HLAminer was used to predict the main HLA-I (Table 1) and HLA-II (Table 2) alleles from three ONT and one PacBio HiFi human WGS datasets corresponding to three individuals (NA12878, NA19240 and NA24385), directly from streamed minimap2 sequence alignments. Group-level (2-digit) predictions matched all but one (94\% accuracy) and three (88\% accuracy) clinically-derived types for HLA-I and -II, respectively (Tables 1 and 2). Allele-level (4-digit) predictions were largely consistent between HLAminer and clinically determined types, for both HLA-I and -II, and considering the low sequencing coverage (e.g. 10X) and low-base accuracy of earlier base-called dataset versions (guppy v1.4.0), performed robustly well. Interestingly, when compared to HLA-II, HLA-I allele predictions from higher-coverage (39X and 67X) promethION datasets were closer to their clinically typed counterparts, and in one case is even corroborated between two different sequencing platforms (HLA-B*35:01 predicted from both ONT and PacBio vs. clinically typed as B*35:08, for NA24385/HG002). The run time for predictions varied between datasets, and roughly scaled with the number of sequencing reads to align, with minimap2 being the primary pipeline bottleneck for both run time and random-access memory (Table 3).\par

\begin{table}[h!]
\hrule \vspace{0.1cm}
\caption{\label{tab:ig}HLA-I predictions from HLAminer (A,B,C,D) directly from human whole genome shotgun long sequencing reads datasets, compared to their clinically determined HLA types (bold). The lowest numbered allele is indicated when two or more alleles are predicted for a group}
\centering
\begin{tabular}{lccccc}
\toprule
\textbf{Dataset} & \textbf{HLA-A} & \textbf{HLA-B} & \textbf{HLA-C}  \\
\midrule
\textbf{NA12878, clinical} &      \textbf{01:01/11:01} & \textbf{08:01/56:01} &     \textbf{01:02/07:01}  \\
 (A) PromethION, SRR10965087, 39X &      01/11:01P  &      08:01P/56:01P &     01:02P/07:01P   \\
 \toprule
 \textbf{NA19240, clinical} &      \textbf{30:01/68:02} & \textbf{35:01/57:03} &     \textbf{04:01/18:01}   \\
 (B) PromethION, ERR2585115, guppy1.4.0, 10X &      31:16/68:02P  &      35:01P/57:01P &     04:01P/18:01P \\
 \toprule
 \textbf{NA24385, clinical} &      \textbf{01:01/26:01} & \textbf{35:08/38:01} &     \textbf{04:01/12:03}  \\
 (C) PromethION, guppy5.0.6, R9.4.1, 67X &      01:01P/26:01P &      35:01P/38:01P &     04:01P/12:03P \\
 (D) PacBio Sequel CCS SRX5327410, 30X &   01:01P/26    &  35:01P/38:01P     &  04:65P/12:03P   \\

\bottomrule
\end{tabular}
\end{table}

\begin{table}[h!]
\hrule \vspace{0.1cm}
\caption{\label{tab:ig}HLA-II predictions from HLAminer (A,B,C,D) directly from human whole genome shotgun long sequencing reads datasets, compared to their clinically determined HLA types (bold). The lowest numbered allele is indicated when two or more alleles are predicted for a group}
\centering
\begin{tabular}{lccccccc}
\toprule
\textbf{Dataset} & \textbf{DQA1} & \textbf{DQB1} & \textbf{DRB1} & \textbf{DPA1} & \textbf{DPB1} \\
\midrule
\textbf{NA12878, clinical} &      \textbf{01:01/05:01} & \textbf{02:01/05:01} &     \textbf{01:01/03:01}  &  \textbf{01:03/02:01}  &   \textbf{04:01/14:01}  & \\
 (A)  &      01:01P/05:01P  &      02:01P/03:01P &     01:02P/03:01P & 01:03P/02:01P & 91:01P/14:01P   \\
 \toprule
 \textbf{NA19240, clinical} &      \textbf{01:02/05:01} & \textbf{05:02/03:01} &     \textbf{16:02/12:01} & \textbf{02:01/02:02} &     \textbf{01:01/01:01}  \\
 (B)  &      01:02/05:01  &      05:02P/03:01P &     16:02P/12:01P & 02:01P/02:02P & 90:01P/01:01P \\
 \toprule
 \textbf{NA24385, clinical} &      \textbf{01:01/03:01} & \textbf{05:01/03:02} &     \textbf{10:01/04:02}  & \textbf{NA} & \textbf{NA}\\
 (C)  &      01:01P/03:01P &      05:01P/03:01P &     10:01P/04:03P & 01:03P/02:01P & 04:01P/23:01P \\
 (D)  &   01:01P/03:01P &      05:01P/03:02P &     10:01P/04:03P & 01:03P/- & 04:01P/666:01   \\

\bottomrule
\end{tabular}
\end{table}

\begin{table}[h!]
\hrule \vspace{0.1cm}
\caption{\label{tab:ig}HLAminer resource usage (48 threads, Intel(R) Xeon(R) Gold 6150 CPU @ 2.70GHz)}
\centering
\begin{tabular}{lcc}
\toprule
\textbf{HLAminer run} & \textbf{Wall clock (h:mm)} & \textbf{Peak memory (GB)} \\
(A) & 2:20 & 44.7\\
(B) & 0:38 & 40.2\\
(C) & 6:21& 41.4 \\
(D) & 1:20& 42.1 \\
\bottomrule
\end{tabular}
\end{table}

\section{Conclusions}
I present a simple protocol for predicting the likely HLA makeup of human samples, by streaming associated WGS (or RNA-seq) long sequencing read alignments directly into HLAminer. The method is as easy as running minimap2 itself, is agnostic to sequence aligners as long a they output alignments in the sam format, and is robust to relatively high base error and low sequence coverage. As more and more genomics projects include a long sequencing read data component, I expect HLAminer to continue providing valuable HLA predictions and insights to the scientific community.\par

\bibliographystyle{unsrt}  


%

\end{document}